\documentclass[aps,prl,twocolumn,showpacs,amsmath,amssymb,floatfix,superscriptaddress]{revtex4}
\usepackage{graphicx}

\begin{document}

\title{A large-deviation approach to space-time chaos}
\date{\today}

\author{Pavel V. Kuptsov}
\email{p.kuptsov@rambler.ru}
\affiliation{Department of Technical Cybernetics and Informatics,
  Saratov State Technical University, Politekhnicheskaya 77, Saratov
  410054, Russia}

\author{Antonio Politi}
\affiliation{CNR - Istituto dei Sistemi Complessi, via Madonna del Piano 10,
I-50019 Sesto Fiorentino, Italy}
\affiliation{Institute for Complex Systems and Mathematical Biology, SUPA,
University of Aberdeen, Aberdeen AB24 3UE, United Kingdom}

\begin{abstract}
In this Letter we show that the analysis of Lyapunov-exponents fluctuations
contributes to deepen our understanding of high-dimensional chaos. 
This is achieved by introducing a Gaussian approximation for the large
deviation function that quantifies the fluctuation probability. More precisely,
a diffusion matrix $\bf D$ (a dynamical invariant itself) is measured
and analysed in terms of its principal components. The application of this
method to three (conservative, as well as dissipative) models, allows: (i) 
quantifying the strength of the effective interactions among the different 
degrees of freedom; (ii) unveiling microscopic constraints such as those
associated to a symplectic structure; (iii) checking the hyperbolicity of the
dynamics.
\end{abstract}



\maketitle

\textit{Introduction -} There are two complementary reasons to
investigate the links between statistical mechanics and space-time
chaos. On the one hand, (equilibrium) statistical mechanics provides
an effective framework to describe the evolution of nonlinear
systems. This is achieved through the introduction of the so-called
thermodynamic formalism \cite{ruelle} and is based on a suitable
partition of the phase-space and the consequent interpretation of the
time-axis as an additional spatial direction. This approach proved
to be very effective in the characterization of low-dimensional
systems and has contributed to establish, e.g., the relationship
between Lyapunov exponents on the one side and fractal dimension or
the Kolmogorov-Sinai entropy, on the other \cite{book}. A
generalization of the approach to spatially extended systems is
formally possible, but almost unfeasible, because of the difficulty to
construct appropriate phase-space partitions \cite{politi}. On the other hand, a
detailed understanding of high-dimensional chaos can help bridging the
gap between microscopic and macroscopic evolution, thereby laying the
foundations for a dynamical theory of (non)equilibrium statistical mechanics.
In this perspective, the estimation of a suitable large deviation function
appears to be the most promising strategy. This idea proved already fruitful in
the context of a stochastic dynamics, where some exact calculations have been
performed in simple but non-trivial models of interacting
particles~\cite{derrida,new}. In the context of chaotic systems, instead, this
approach is the core of the Gallavotti-Cohen fluctuation
theorem~\cite{gallavotti}, that is proved under the hypothesis that, in the
thermodynamic limit, the evolution of typical dynamical systems is effectively
hyperbolic.

In this Letter, we propose an approach that can contribute to make
progress along both directions, without introducing any assumption of the
underlying dynamics. More precisely, we suggest to study the fluctuations
of the Lyapunov exponents (LEs) along the lines of the multifractal
theory \cite{book}. One of the advantages of dealing with LEs and their
fluctuations (in the long-time limit) is that they are dynamical invariants,
i.e. they are independent of the parametrization of the phase space.
An exact implementation in generic nonlinear models is out of question.
Nevertheless, here we show that useful information can be extracted by working
within the Gaussian approximation. For instance, we show that the
(cross)correlations among all pairs of LEs and, in particular, their scaling
behavior with the system size allows estimating the strength of the effective
interactions that spontaneously emerge among the various degrees of freedom.
Notice that our analysis goes beyond the usual extensivity assessment of
space-time chaos, that is linked to the existence of a limit Lyapunov spectrum.
In fact, we will see that the fluctuations of a chain of contiguous
non-interacting systems are substantially different from those of
a typical chain of interacting systems.
Finally, our approach allows testing the hyperbolicity
of the underlying dynamics, by: (i) comparing the results obtained for different
defintions of the Lyapunov exponents, (ii) testing phenomena like the dominance
of Oseledec splitting \cite{dos}, and (iii) quantifying dimension variability
\cite{lai}.

\textit{Theory -} Let $\Lambda_i(\tau)$ denote the $i$th expansion factor over a
time $\tau$ in tangent space. The rate $\lambda_i = \Lambda_i(\tau)/\tau$ is the
so-called finite-time Lyapunov exponent (FTLE), which, in the infinite-time
limit, converges to the LE $\overline \lambda_i$ (here and in the following,
overlines denote time averages). For finite $\tau$, FTLEs fluctuate around the
asymptotic values. The theory of large deviations suggests that, in the
long-time limit, the probability distribution $P(\boldsymbol{\lambda},\tau)$ 
(where $\boldsymbol{\lambda} = \{\lambda_1,\lambda_2,\ldots,\lambda_N\}$ and $N$
is the number of degrees of freedom) scales with $\tau$ as
\begin{equation}
P(\boldsymbol{\lambda},\tau) \underset{\tau\to\infty}{\propto}
{\rm e}^{-S(\boldsymbol{\lambda})\tau} 
\label{eq:ld}
\end{equation}
where $S(\boldsymbol{\lambda})$ is the positive-definite large deviation
function
whose minimum (equal to zero) is achieved in correspondence of the
LEs $\overline \lambda_i$. $S$ has been mostly studied in contexts
where $\boldsymbol{\lambda}$ reduces to a scalar variable $\lambda$, as it
happens for low-dimensional chaos, where it is known that $S(\lambda)$
is itself a dynamical invariant~\cite{book}. This is because FTLE fluctuations
originate from passages of a chaotic trajectory in the vicinity of periodic
orbits with different stability properties. There is no reason to doubt that
dynamical invariance is lost upon increasing the dimensionality of
the phase-space.

If a system is the Cartesian product of uncoupled variables, $S$ is
the sum of functions, each dependent on a single $\lambda_i$, but interactions
bring new terms. Although determining $S$ is too ambitious a task, 
relevant features can be uncovered by expanding it around the minimum
$\overline{\boldsymbol{\lambda}}$, and retaining the first (quadratic) non-zero
term,
\begin{equation}
S(\boldsymbol{\lambda}) \approx \frac{1}{2}
(\boldsymbol{\lambda}-\overline{\boldsymbol{\lambda}}){\bf Q}
(\boldsymbol{\lambda}-\overline{\boldsymbol{\lambda}})^\dag 
\label{eq:gauss}
\end{equation}
($\dag$ denotes the transpose), an approximation that is equivalent to
assuming a Gaussian distribution. In practice, it is preferable to consider the
symmetric matrix ${\bf D} = {\bf Q}^{-1}$. In fact, the elements $D_{ij}$ can be
directly determined by estimating the (linear) growth rate of the
(co)variances of $(\Lambda_i(\tau)- \overline \lambda_i \tau)$,
\begin{equation}
D_{ij} = \lim_{\tau\to\infty}
\left( \overline{\Lambda_i(\tau)\Lambda_j(\tau)}-
\overline \lambda_i \overline \lambda_j \tau^2 \right)
/\tau .
\label{eq:diff}
\end{equation}
There are three basic definitions of FTLEs. One can compute them:({\it i}) by
repeatedly applying the Gram-Schmidt orthogonalization procedure to a set of
linearly independent perturbations (backward or Gram-Schmidt Lyapunov vectors);
({\it ii}) by performing this procedure along the negative time axis (forward
Lyapunov vectors); ({\it iii}) by making reference to the covariant Lyapunov
vectors \cite{ginelli}. In the infinite-time limit, the three methods produce
identical LEs. For finite but long times, as long as FTLE fluctuations are
connected to ``visits" of different periodic orbits, the three defintions should
be again equivalent. In fact, we have systematically verified that the
correlations computed with the methods ({\it i}) and ({\it iii}) are basically
indistinguishable as soon as the diffusive asymptotic behavior sets in 
\cite{noteme}. This result provides a first evidence of the effective
hyperbolicity of the underlying dynamics. In fact, it implies that the
differences induced by the presence of homoclinic tangencies are so rare that
they do not affect our perturbative analysis.

The information contained in $\bf D$ can be expressed in a compact form by
determining its (positive) eigenvalues $\mu_k$ ($k=1,N$), which represent the
fluctuation amplitudes along the most prominent directions. Some of the
eigenvalues may turn out to vanish because of more or less hidden constraints.
For instance, in the presence of a constant phase-space contraction rate,
$\sum_i \lambda_i = {\rm const}$ and all $\{\lambda_i\}$ $n$-tuples lie in a
same hyperplane. As a result, one eigenvalue of $\bf D$ is equal to zero: its
corresponding eigenvector is perpendicular to the hyperplane itself. Another
instructive case is that of symplectic dynamics: since the LEs come in pairs
whose sum is zero, the fluctuations of the negative LEs are perfectly
anticorrelated with those of the positive ones, so that $D_{ij}$ has an
additional symmetry $\cal S$, i.e., $D_{N+1-i,j}=D_{i,N+1-j} =-D_{ij}$. 
Altogether, the possible existence of zero eigenvalues reinforces the choice of
studying $\bf D$ rather than its ill-defined inverse $\bf Q$. Moreover, since
the matrices $\bf Q$ and $\bf D$ are diagonal in the same basis, and the
eigvenvalues of $\bf Q$ are the inverse of those of $\bf D$, we can infer the
scaling behavior of the former ones from that of the latter. One must simply be
careful and discard the {\it redundant} variables, associated to the zero
eigenvalues. In particular, since, as we shall see, $\mu_k \propto 1/N$, the
large deviation function $S$ turns out to be proportional to the number $N$ of
degrees of freedom, i.e., it is an extensive quantity.

\begin{figure}[!ht]
\includegraphics[clip=true,width=0.4\textwidth]{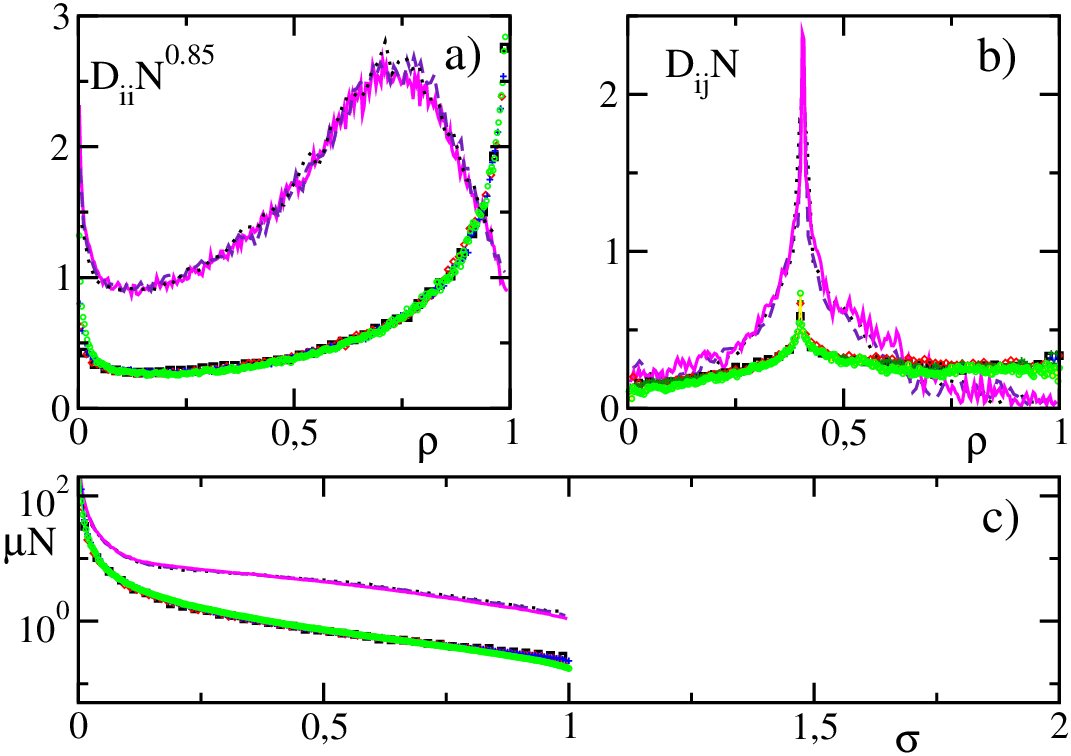}
\caption{(Color online) Diffusion coefficients in a chain of H\'enon (symbols)
and symplectic (lines) maps. In all panels, squares,
diamonds, plusses  and circles refer to $N=40$, 80, 160, and 320 respectively, 
while dotted, dashed and solid lines correspond to $N=32$, 64, and 128,
respectively. The results have been obtained by iterating the chain over
$5\times 10^6$ time steps. Panel a) contains the diagonal elements,
$\rho=(i-1/2)/N$; panel b) refers to the column
$j=2N/5$, $\rho=i/N$; panel c) refers to the eigenvalues of $\bf D$, ordered
from the largest to the smallest one, $\sigma = k/N$.}
\label{fig1}
\end{figure}

\textit{Model analysis -} We start the numerical analysis by studying
a chain of H\'enon maps \cite{torc92}
\begin{equation}
x_n(t+1)=a- [x_n(t)+ \varepsilon {\cal D} x_n(t)]^2 +  bx_n(t-1)  \, ,
\label{eq:henon}
\end{equation}
where ${\cal D} x_n \equiv (x_{n-1} - 2 x_n + x_{n+1})$ is the
discrete Laplacian operator. We have chosen $a=1.4$, $b=0.3$,
$\varepsilon=0.025$ and used periodic boundary conditions (the same
conditions have been chosen in the other models too).  The results are
shown in Fig.~\ref{fig1}. In panel a) we report the self-diffusion
coefficients $D_{ii}$ (see the symbols). The clean overlap of the scaled curves
indicates that $D_{ii}(\rho) \approx 1/N^{0.85}$. This means that the LEs
self-average in the thermodynamic limit. The singular behavior exhibited by
$D_{ii}(\rho)$  for $\rho \to 1$ follows from the different scaling behavior of
the first and $N$th exponent which decrease as $1/\sqrt{N}$.
In Fig.~\ref{fig1}b we plot $D_{ij}$ along the column $j=2N/5$. The off-diagonal
terms decrease as $1/N$, so that the matrix $\bf D$ becomes increasingly
diagonal in the thermodynamic limit. Finally, the eigenvalue spectrum $\mu_k$
(see Fig.~\ref{fig1}c) decreases like $1/N$. This implies that the eigenvalues
of $\bf Q$ are proportional to $N$, i.e. the large deviation function is an
extensive observable. Moreover, the $1/\sigma$ singularity at $\sigma=0$ means
that the leading eigenvalue $\mu_1$ does not decrease, i.e., there exists one
direction in phase-space along which fluctuations survive even in the
thermodynamic limit. The physical meaning of this feature is to be understood.
Finally, the eigenvalue spectrum exhibits a remarkable and unexpected property:
half of it is equal to zero. A close inspection of the whole correlation matrix
reveals that this is because $\bf D$ is $\cal S$-symmetric. By further
investigating the Jacobian matrix $\bf J$, we have discovered that it indeed
satisfies the symplectic-like condition ${\bf J}{\bf A}{\bf J}^{\rm T}=-b{\bf
A}$ (see \cite{unpu}). Unlike the similar case studied in Ref.~\cite{dressler},
here ${\bf A}$ is a generic antisymmetric matrix depending on $t$. Altogether,
these results indicate that LEs come in pairs, such that
$\overline\lambda_i+\overline\lambda_{N+1-i}=\ln b$. 

Next we have studied a chain of symplectic maps,
\begin{eqnarray}
p_n(t+1) &=& p_n(t) + K(\sin \Delta\theta_{n+1}(t)
-\sin \Delta \theta_n(t)) \nonumber \\
\theta_n(t+1) &=& \theta_n(t) + p_n(t+1)
\end{eqnarray}
where both $p_n$ and $\theta_n$ are defined modulus $2\pi$ and $\Delta
\theta_n = \theta_n-\theta_{n-1}$. The model has been simulated for
$K=4$. In Fig.~\ref{fig1} (see the lines), one can notice that the overall
scenario is very similar to that one observed in the chain of H\'enon maps,
including the behavior of the diagonal elements. The major difference concerns
$D_{NN}$ which, instead of decreasing faster, it now decreases slower
than in the bulk.

\begin{figure}[!ht]
\includegraphics[clip=true,width=0.4\textwidth]{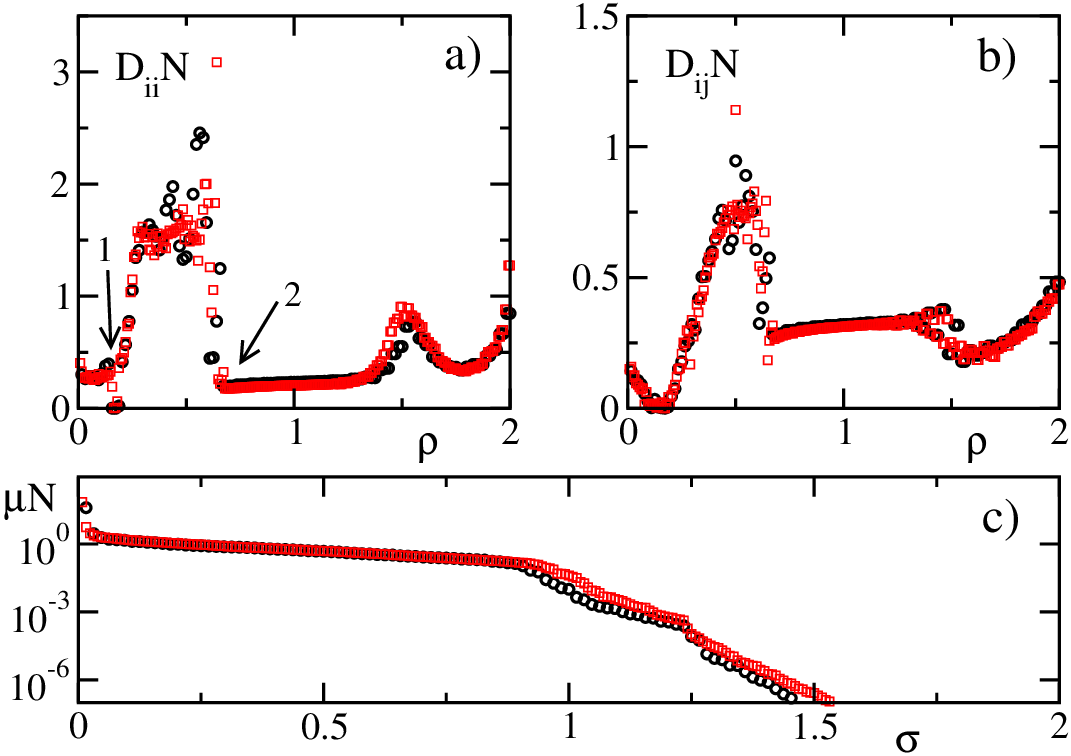}
\caption{(Color online) Diffusion coefficients in a chain of Stuart-Landau
oscillators. In all panels circles and squares refer to $N=64$ and
128, respectively. The panels are organized in the same way as in the
first figure. The column reported in panel b) corresponds to $j=N/2$.}
\label{fig3}
\end{figure}

Finally, we have considered a chain of Stuart-Landau oscillators as an
example of a continuous-time dissipative system. The model can be viewed as
the spatial discretization of a complex Ginzburg-Landau equation,
a prototypical model of space-time chaos. The evolution equation
writes
\begin{equation}
\dot a_n = a_n -(1+ic)|a_n|^2a_n+(1+ib)h^{-2} {\cal D}a_n \, .
\end{equation}
We have fixed $c=3$, $b=-2$ and $h=1/2$, which corresponds to a regime
of amplitude turbulence \cite{parlitz}. In this model we cannot draw
clear conclusions on the scaling behavior of the $\bf D$ elements, because of
larger finite-size corrections (see Fig.~\ref{fig3}). However, the eigenvalues
behave quite similarly to the two previous cases: (i) the overall spectrum
scales as $1/N$; (ii) the maximum eigenvalue remains finite for increasing $N$;
(iii) a large fraction of the spectrum is nearly equal to zero. In this case,
the singularity is due to the appearance (beyond a certain $\rho$-value)
of pairs of degenerate LEs~\cite{parlitz} which fluctuate synchronously.
Notice also the drops of the diffusion coefficient indicated by arrows 1
and 2 in Fig.~\ref{fig3}a that are discussed below.

\textit{Discussion - } The common property exhibited by all of the three
models is the $1/N$ scaling of the eigenvalue spectrum
of the diffusion matrix $\bf D$. This implies that the matrix $\bf Q$ and,
thereby, the large deviation function $S$ are proportional to the number of
degrees of freedom,
i.e., $S$ is an {\it extensive} observable. It is interesting to notice that such
a property holds in spite of the long-range correlations that
are revealed by the strength of the off-diagonal terms of $\bf D$ (in all
models, they provide a substantial contribution to the scaling behavior
of the eigenvalues). 

\begin{figure}[!ht]
\includegraphics[clip=true,width=0.4\textwidth]{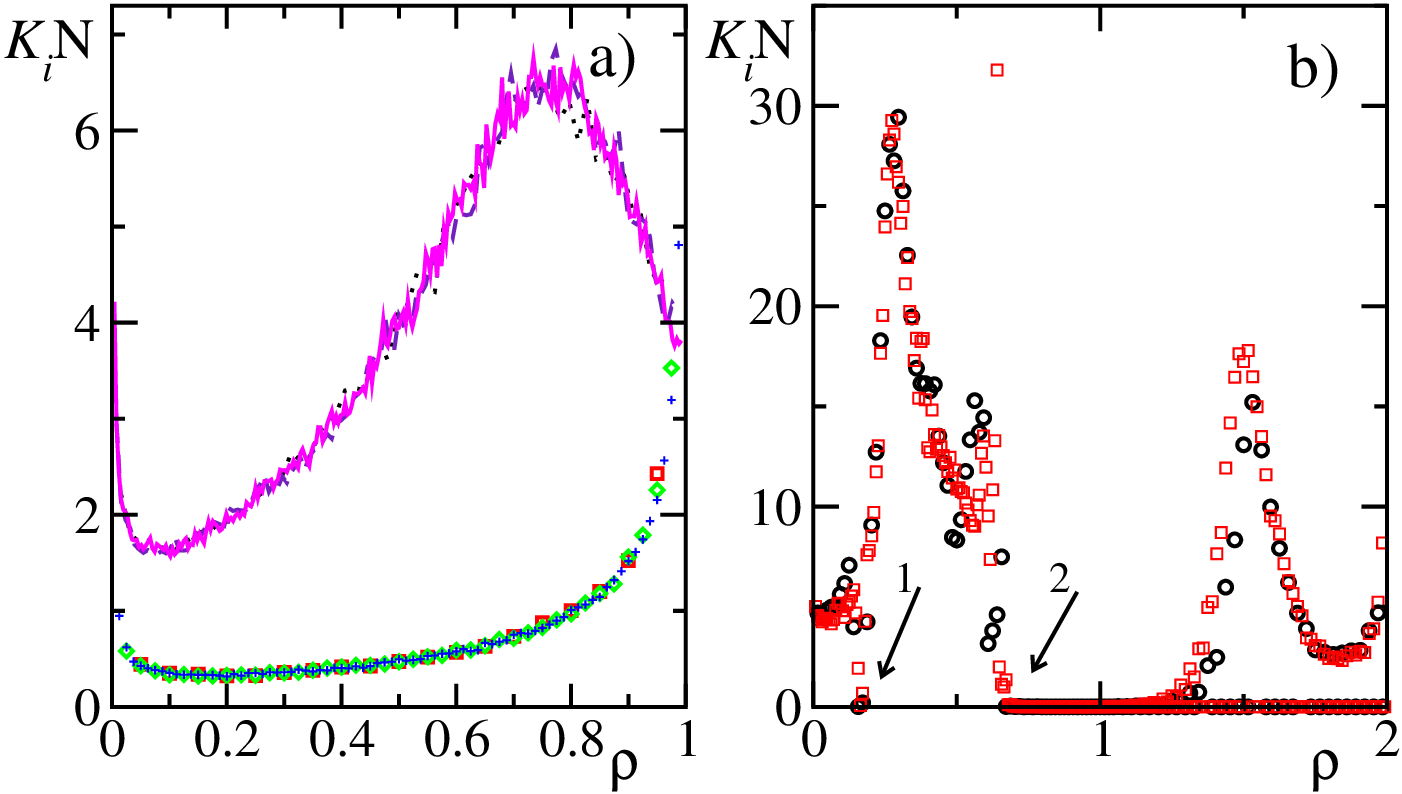}
\caption{(Color online) Rescaled diffusion coefficients $K_i$ of the
  LE differences $\delta \lambda$ in H\'enon (symbols) and symplectic
  (lines) maps (panel a) and Stuart-Landau oscillators (panel b), $\rho=i/N$.
  Same notations as in the two previous curves.}
\label{fig5}
\end{figure}

Next we discuss some physical implications of the structure of the large
deviation function $S$. We start from the occurrence of occasional
changes in the order of the FTLEs, a feature that is related to the
concept of dominated Oseledec splitting \cite{dos}. The splitting is
dominated with index $i$ if there exists a {\it finite} $\tau_0$ such that
$\lambda_i>\lambda_{i+1}$ for all $\tau>\tau_0$ \cite{yang}. This
property implies the absence of tangencies between the corresponding
Oseledec subspaces identified by the $i$th vector and the subsequent one
\cite{dos}. The probability of order-exchanges can be inferred from
the fluctuations of $\delta \lambda_i= \lambda_i-\lambda_{i+1}$. Its diffusion
coefficient $K_i$ can be expressed in terms of the ${\bf D}$ elements,
$K_i = D_{ii}+D_{i+1,i+1}-2D_{i,i+1}$. Since the probability of an exchange
of FTLEs is equal to the probability $P_i^w$ of observing a negative
$\delta \lambda_i$, we have, in the Gaussian approximation,
\begin{equation}
P_i^w \approx \frac{1}{2} {\rm erfc} \left \{ (\overline \lambda_i -\overline
\lambda_{i+1}) \sqrt{\frac{\tau}{2K_i}} \right\} ,
\end{equation}
where ``$\rm erfc$" is the complementary error function. The analysis
of the three models reveals that, in the bulk, $K_i$ scales always as
$1/N$ (see Fig.~\ref{fig5}) \cite{notescl}.  
Since the distance between consecutive LEs scales also as $1/N$ (this follows
from the very existence of a limit LE spectrum), one can conclude that
$P_i^w \approx {\rm erfc}(c_i\sqrt{\tau/N})/2$ for some $c_i\geq 0$. 
This means that in the large $N$ limit, order exchanges occur with a finite
probability and no dominated splitting is present. However, in the
H\'enon maps, at $\rho=1$, there is a gap in the Lyapunov spectrum.
Therefore, since $K_N$ vanishes (as $1/\sqrt{N}$), the probability of
order exchanges goes to zero, indicating that stable and unstable manifold are
mutually transversal and the system effectively hyperbolic. The
absence of a gap in the Lyapunov spectrum of the symplectic maps prevents us
from drawing a similar conclusion in that model. In the Stuart-Landau chain,
$P_i^w$ vanishes close to arrow ``2" (see Figs.~\ref{fig3}a and \ref{fig5}b), 
since $K_i=0$ (and $\overline \lambda_i\neq\overline \lambda_{i+1}$), thus
implying that the splitting is dominated~\cite{note1}. This suggests the
existence of two transversal subspaces, consistently with the
claim that the attractor is embedded in a supporting manifold containing
the physical modes \cite{yang,parlitz}. Since the dimension of the
supporting manifold is even larger than the Kaplan-Yorke dimension
(equal to $0.27N$), we must conclude that the overall dynamics is not
hyperbolic.

Now, we analyse the invariant measure, introducing the expansion rate
$\mathcal{L}(\mathcal{D}) = \sum_i^\mathcal{D} \lambda_i$ of a generic
volume of dimension $\mathcal{D}$, over a time $\tau$. The Kaplan-Yorke
dimension $\mathcal{D}_f$ is obtained by imposing
$\overline{\mathcal{L}}(\mathcal{D}_f)=0$ \cite{notefr}. Under the assumption of
small fluctuations, one can express the diffusion
coefficient $\Delta_\tau^\mathcal{D}$ of $\mathcal{D}$ in terms of the
analogous coefficient
$\Delta^\mathcal{L}= \sum_{i,j}^\mathcal{D}D_{ij}$ of $\mathcal{L}$,
by linearizing the function $\overline{\mathcal{L}}(\mathcal{D})$ around
$\mathcal{D}_f$. This leads to
$\Delta_\tau^\mathcal{D}= \Delta^\mathcal{L}/
({\overline  \lambda}_{\mathcal{D}_f})^2$.
The dimension fluctuations $\Delta_\varepsilon^\mathcal{D}$, can now be
estimated by invoking an Ansatz similar to Eq.~(\ref{eq:ld}) which, in the
Gaussian approximation, writes
$P(\mathcal{D})\propto \varepsilon^{(\mathcal{D}-\mathcal{D}_f)^2/
(2\Delta_\varepsilon^\mathcal{D})}$, where the box-size $\varepsilon$ must be
linked
to the time variable. By following Ref.~\cite{grass}, it is natural to assume
that $\varepsilon \approx \exp(-|\overline{\lambda}_{\mathcal{D}_f}|\tau)$,
thereby obtaining
$\Delta_\varepsilon^\mathcal{D}=
\Delta_\tau^\mathcal{D} \overline{\lambda}_{\mathcal{D}_f} =
\Delta^\mathcal{L}/\bar \lambda_{\mathcal{D}_f}$.
In the chain of H\'enon maps,
$\Delta_\varepsilon^\mathcal{D} \approx 0.12 N$, i.e. dimension fluctuations are
extensive. This implies that the na\"ive idea is wrong and it is
necessary to build a more refined picture to refer to high-dimensional chaotic
attractors.

\textit{Conclusions} We have shown that a fluctuation analysis can deepen our
understanding of high-dimensional chaos. The main result is the discovery of a
subtle form of extensivity, i.e. the proportionality of the large deviation
function to the system size. This result is nontrivial, since it arises in a
context of effective long-range correlations and there are even examples
of stochastic models, where the large deviation function is not extensive
\cite{new}. As for the discrepancy between the scaling exponent of the
diagonal elements and of the eigenvalues of $\bf D$ (0.85 vs. 1) it
is necessary to study larger sizes to decide whether it is due to
finite-size corrections. 
Moreover, our approach provides a new way of
investigating the  hyperbolicity of a given dynamics (including dimension
variability), although we are aware that the last word can be said only
by going beyond the perturbative approach described in this Letter. The method
introduced in Ref.~\cite{kurchan} to identify trajectories with unprobable
stability properties, makes this perspective not so remote.

\textit{Acknolwedgements} We thank S. Lepri, A. Pikovsky, H. Chat\'e and
K. Takeuchi for useful discussions.


\begin{thebibliography}{xx}
\bibitem{notescl} The divergence visible in Fig.~\ref{fig1}(b) close to the diagonal
implies that $D_{i,i+1}$ scales as $1/N^{0.85}$: this term is large enough
to cancel the diagonal $1/N^{0.85}$ contribution, thereby allowing
$K_i$ to scale as $1/N$.
\bibitem{note1} The same is not true close to ``1", since the
vanishing of $K_i$ is accompanied by the presence of identical zero LEs.
\end{thebibliography}
\end{document}